\documentclass[%
aip,
pop,
amsmath,
amssymb,
reprint,
]{revtex4-2}

\usepackage{xcolor}
\usepackage{graphicx}
\usepackage{dcolumn}
\usepackage{bm}
\usepackage[mathlines]{lineno}

\begin{document}

\preprint{AIP/123-QED}

\title{
    Ion Temperature Anisotropy Limits from Magnetic Curvature Scattering in Magnetotail Reconnection Jets
}%

\author{Louis Richard}
\email{louis.richard@irfu.se}
\affiliation{
    Swedish Institute of Space Physics, Uppsala 751 21, Sweden
}
\affiliation{
    Department of Earth, Planetary, and Space Sciences, University of California, Los Angeles, California, 90095, USA
}

\author{Anton V. Artemyev}
\affiliation{
    Department of Earth, Planetary, and Space Sciences, University of California, Los Angeles, California, 90095, USA
}

\author{Cecilia Norgren}
\affiliation{
    Swedish Institute of Space Physics, Uppsala 751 21, Sweden
}
\affiliation{Department of Physics and Engineering, University of Bergen, Bergen 5007, Norway}

\author{Xin An}
\affiliation{
    Department of Earth, Planetary, and Space Sciences, University of California, Los Angeles, California, 90095, USA
}

\author{Sergey R. Kamaletdinov}
\affiliation{
    Department of Earth, Planetary, and Space Sciences, University of California, Los Angeles, California, 90095, USA
}

\author{Yuri V. Khotyaintsev}
\affiliation{
    Swedish Institute of Space Physics, Uppsala 751 21, Sweden
}
\affiliation{
    Department of Physics and Astronomy, Uppsala University, Uppsala 751 20, Sweden
}

\date{\today}

\begin{abstract}
    In collisionless plasmas, relaxation of the deviations of ion velocity distribution functions (VDFs) from local thermodynamic equilibrium occurs through particle interactions with electromagnetic fields. 
    In particular, in the Earth's magnetotail, the deviations of the ion VDFs, typically consisting of multiple components, from the equilibrium must be limited to maintain stability of the current sheet. 
    Curvature scattering is a leading candidate mechanism to limit such deviations, but its role remains insufficiently understood. 
    We investigate the limits of ion temperature anisotropy in a magnetotail-like configuration by modeling a quasi-1D current sheet with a finite magnetic field curvature and three ion populations. 
    We derive analytical thresholds for anisotropy based on current sheet stability and validate against spacecraft observations and numerical simulations. 
    Our findings demonstrate that curvature scattering imposes limits on ion anisotropies, thereby maintaining the stability of the current sheet.
\end{abstract}


\maketitle

\section{Introduction}
Magnetic reconnection is a fundamental plasma process that explosively converts magnetic energy into particle kinetic and internal energy by reconfiguring the magnetic field topology. 
Magnetic reconnection powers energetic events in astrophysical systems such as black hole flares, accretion disks, and solar flares, plays a key role in magnetotail dynamics, and limits confinement in fusion plasmas~\cite{zweibel_magnetic_2009,yamada_magnetic_2010}. 
Extensive \textit{in situ} spacecraft observations and numerical simulations demonstrated that multiple mechanisms conspire to energize charged particles (ions and electrons) during anti-parallel magnetic reconnection  ($B_g=0$), as typically observed in the Earth's magnetotail~\cite{norgren_electron_2025,oka_particle_2023}. 
In particular, ion energization primarily results from acceleration by the Hall and reconnection electric fields. 
In the reconnection inflow, ions are approximately in a local thermodynamic equilibrium characterized by a Maxwell-Boltzmann velocity distribution function (VDF). 
The Hall electric field associated with the quadrupolar magnetic field arising from electrons flowing toward and away from the X-line~\cite{uzdensky_physical_2006}, ballistically accelerates ions from the reconnection inflow towards the center of the reconnecting current sheet, forming two cold beams counter-streaming along the normal magnetic field with parallel, i.e., magnetic field aligned, anisotropy~\cite{wygant_cluster_2005}. 
Alternatively, similar counter-streaming ion beams can form through pick-up acceleration, in which ions entering the reconnection outflow are swept up and accelerated by the bulk flow~\cite{drake_ion_2009,eastwood_ion_2015,liu_ionbeamdriven_2019}. 
Meanwhile, the reconnecting electric field accelerates demagnetized ions that undergo quasi-adiabatic transient Speiser orbits, characterized by a drifting, bean-shaped ion VDFs at the center of the current sheet~\cite{cowley_current_1983,burkhart_particle_1992}. 
Both mechanisms generate highly anisotropic ion VDFs that carry a substantial fraction of the current density.  

\textit{In situ} observations in the Earth's magnetotail demonstrated that the anisotropy of the ion VDFs is efficiently limited downstream in the reconnection outflow~\cite{wu_proton_2013,richard_fast_2023}. 
Typically, in collisionless plasmas, this relaxation is thought to be achieved through wave-particle interaction with electromagnetic waves excited by temperature anisotropy-driven instabilities~\cite{gary_theory_1993}. 
In contrast, within the reconnection outflow, numerical simulations and \textit{in situ} observations indicate that the linear growth time of temperature–anisotropy–driven instabilities is small compared with the time it takes for the outflow to propagate across distances comparable to the system size~\cite{hietala_ion_2015,richard_fast_2023}. 
Consequently, the excited waves cannot reach large amplitudes. 
Therefore, pitch-angle scattering by wave-particle interaction cannot explain the fast relaxation of the ion VDFs into a lower energy state. 
Instead, \textit{in situ} observations suggested that curvature scattering can provide such efficient relaxation~\cite{richard_fast_2023}.

Curvature scattering is due to the chaotization of the ion motion caused by nonlinear resonance between the particle's bounce and gyromotion in the curved magnetic field~\cite{birmingham_pitch_1984,buchner_regular_1989}. 
This mechanism has been extensively studied theoretically in simple magnetotail-like quasi-1D current sheet configurations with a finite normal magnetic field component and no guide field (see Ref.~\onlinecite{zelenyi_quasiadiabatic_2013} and references therein). 
In such current sheet configurations, the ion dynamics is controlled by the curvature parameter $\kappa = \sqrt{R_c / r_i}$, where $R_c = 1 / |\bm{\hat{b}}\cdot \nabla \bm{\hat{b}}|$, with $\bm{\hat{b}} = \bm{B} / B$, where $\bm{B}$ is the magnetic field vector and $B$ its magnitude, and $r_i = v_{i\bot} / \omega_{ci}$ the ion gyroradius, with $\omega_{ci}=eB/m_i$ the ion gyrofrequency. 
When $\kappa \gg 1$, the magnetic moment $\mu= m_iv_\bot^2 / 2B$ is conserved with an exponential accuracy, and the ion motion is adiabatic. 
However, as $\kappa \rightarrow 1$, the ion motion becomes weakly chaotic and leads to a weak diffusion with $\Delta \mu / \mu \sim \exp (-2\kappa^2 / 3)$. 
At $\kappa = 1$, the ion motion is strongly chaotic with a strong scattering $\Delta \mu / \mu \sim 1$. 
Finally, for $\kappa \ll 1$, the ions follow quasi-adiabatic orbits, e.g., Speiser orbits, experiencing weak scattering $\Delta \mathcal{I}_z/\mathcal{I}_z \sim \kappa$, where $\mathcal{I}_z = \left ( 2\pi\right )^{-1}\oint p_z dz$ is the quasi-adiabatic or current sheet invariant~\cite{buchner_regular_1989}. 
We note that the addition of a guide field and a reconnection electric field introduces rapid geometrical chaos~\cite {artemyev_rapid_2014}.

\textit{In situ} observations in the Earth's magnetotail~\cite{artemyev_ion_2019,kamaletdinov_ion_2025} and numerical simulations of magnetotail reconnection~\cite{hoshino_ion_1998,hietala_ion_2015} have identified three distinct populations in the ion VDFs corresponding to different regimes of motion in the reconnection outflow. 
The first is a hot thermalized isotropic background. 
This population corresponds to the already-thermalized outflow. 
The second population is an adiabatic ($\kappa \gg 1$) population composed of two, cold, counter-streaming, field-aligned beams originating from ions ballistically accelerated by the Hall electric field in the reconnection region~\cite{wygant_cluster_2005}, ions picked up by the outflow~\cite{drake_ion_2009,eastwood_ion_2015,liu_ionbeamdriven_2019}, and ionospheric outflow~\cite{alm_magnetotail_2018,xu_ionospheric_2019,li_quantification_2021}. 
This parallel-anisotropic population contributes to the current density via anisotropic curvature drift. 
The third population consists of supra-thermal ions accelerated by the cross-tail electric field $E_y$, driven by the reconnection electric field and by mesoscale convective structures such as bursty bulk flows~\cite{richard_proton_2022}. 
These ions form a perpendicular, anisotropic, Speiser population, which contributes to the cross-tail current through their quasi-adiabatic motion ($\kappa \ll 1$). 
The interplay between the cold and Speiser anisotropic populations reflects a dynamic equilibrium within the current sheet, in which transient enhancements of anisotropy control the force balance, therefore controlling the stability of the current sheet~\cite{artemyev_ion_2019,an_configuration_2022,arnold_pic_2023}. 
This dynamical evolution of the interplay between ion anisotropy and the stability of the current sheet is an ongoing area of research. 
However, although ion anisotropy is recognized to be necessary to balance the magnetic tension~\cite{rich_balance_1972}, the quantitative bounds of anisotropy that ensure current sheet stability are still not well constrained.

In this paper, we investigate the constraints on ion anisotropy within the current sheet of magnetic reconnection jets in Earth's magnetotail. 
We derive limits on the ion anisotropy in a quasi-1D current sheet with three ion populations based on curvature scattering. 
We show that these limits correspond to thresholds for the stability of the current sheet. 
We validate the model using \textit{in situ} observations in the Earth's magnetotail and numerical simulations.

\section{Theoretical Model}
We consider a quasi-1D planar magnetic field configuration typical of the magnetotail current sheet~\cite{arnold_pic_2023,zhang_radial_2025}, where the reconnecting component of the magnetic field satisfies $\partial_z B_x \simeq \mu_0 j_y$, the normal component is approximately constant ($B_z\simeq \mathrm{const}$), and the curvature radius of the magnetic field is $R_c \simeq B_z/\mu_0 j_y$. 
These current sheets are often polarized by strong electric fields $E_z$ that redistribute ion and electron currents through $\bm{E}\times\bm{B}$ drifts~\cite{hesse_ionscale_1998,lu_hall_2016}. 
However, within the reference frame moving with the $\bm{E}\times\bm{B}$ drift, the electron current is primarily dominated by diamagnetic drifts~\cite{birn_thin_2004} and curvature drifts~\cite{artemyev_contribution_2019}.

\subsection{Current Density and Adiabaticity Parameter}
To estimate the ion contribution to the current density, we model the ion VDFs as consisting of three primary populations as identified in the Earth's magnetotail~\cite{artemyev_ion_2019,kamaletdinov_ion_2025}: (i) a cold ion population ($\kappa > 1$) composed of two field-aligned counter-streaming beams with number density $n_{c}$ and temperature $T_{c} = (T_{c\parallel} + 2T_{c\bot})/3$, with a parallel anisotropy $A_c = T_{c\parallel}/T_{c\bot} > 1$; (ii) a hot, nearly isotropic, population with density $n_{h}$ and temperature $T_{h} \approx T_{h\parallel} \approx T_{h\bot}$; (iii) an energetic Speiser-type population ($\kappa \ll 1$) with density $n_{s}$, temperature $T_{s} = (T_{s\parallel} + 2T_{s\bot})/3$, and perpendicular anisotropy $A_s = T_{s\parallel}/T_{s\bot} < 1$. 
Consequently, cold ions mainly contribute to the parallel temperature anisotropy of the superposed three species population, whereas Speiser ions are responsible for the perpendicular anisotropy and agyrotropy through their quasi-adiabatic, unmagnetized orbits. 
We note that, as we will show later in Sections~\ref{sec:spacecraft-observations} and~\ref{sec:numerical-sim}, the relative contribution of each population can vary across the current sheet~\cite{hoshino_ion_1998,hietala_ion_2015,kamaletdinov_ion_2025}.

The different ion populations contribute to the total current density through different drifts. 
Due to its parallel anisotropy and low thermal energy, the cold ion population primarily contributes via anisotropic curvature drifts~\cite{krall_principles_1973}:
\begin{equation}
    \label{eq:j-cold}
    j_{cy} \approx \dfrac{n_{c}k_BT_{c\bot}(A_{c}-1)}{B^2}\dfrac{\partial B_x}{\partial z} = \dfrac{n_c}{n_i}  (A_c - 1) \eta_c^2 \dfrac{\beta_{i\parallel}}{2} j_y,
\end{equation}
\noindent where $\beta_{i\parallel} = 2 \mu_0 n_i k_B T_{i\parallel} / B^2$ and $\eta_c^2=T_{c\bot}/T_{i\parallel}$. 
When the parallel ion temperature $T_{i\parallel}$ is dominated by the cold-ion population, $\eta_c^2$ quantifies the ratio between their perpendicular thermal speed (set by $T_{c\perp}$) and their parallel bouncing speed (set by $T_{i\parallel}$)\

Due to its large temperature and isotropy, the hot ion population primarily contributes via diamagnetic drifts. 
This gives $j_{hy} \simeq \left [B_x \partial_z P_h  -B_z\partial_x P_h \right ]/B^2$. 
Thus, in current sheets with finite $B_z$, this term reduces to $j_{hy} \simeq -\partial_x P_h/B_z$ around the equatorial plane ($B_x\sim 0$) and becomes significant only in quasi-2D configurations where $\partial_x P_h/B_z$ is comparable to $\partial_z P_h/B_x$~\cite{birn_thin_2004}. 
Although such quasi-2D current sheets can be found in the magnetotail \cite{apatenkov_multispacecraft_2025}, thin, intense current sheets generally show very weak $\partial_x P_h$ gradient, such that $\partial_x P_h / B_z \ll \partial_z P_h / B_x$~\cite{artemyev_configuration_2021,zhang_radial_2025}. 
Therefore, in such thin, intense current sheets, the hot ion contribution to the cross-tail current is negligible $j_{hy}\approx 0$ around the equatorial plane.

Speiser ions contribute to the current density through their quasi-adiabatic trajectories. 
Using an approximation of the Speiser ion VDF~\cite{burkhart_ion_1992,burkhart_particle_1992,artemyev_proton_2010}, the current density from this population can be estimated as $j_{sy} \simeq en_s v_{sD}$, where $v_{sD}$ is the drift speed of the Speiser population. 
Defining a characteristic current $j_0 = en_i v_{iT \bot}$ where $v_{iT \bot} = \sqrt{2k_BT_{i\bot}/m_i}$ is the perpendicular thermal speed, we express the Speiser ion contribution to the current density as $j_{sy} \simeq (n_s/n_i)\eta_s j_0$, with $\eta_s= v_{sD} / v_{iT \bot}$.

Summing these components plus the electron current density $j_{ey} = \eta_e j_0$, where $\eta_e$ is the fraction of electron to characteristic current, the total current density, $j_{y} = j_{ey} + j_{cy} + j_{hy} + j_{sy}$, can be written as
\begin{equation}
    \label{eq:j-tot}
    j_y = j_0 \dfrac{{\eta_e + \dfrac{n_s}{n_i} \eta_s}}{{1 - \dfrac{n_c}{n_i}  (A_c - 1) \eta_c^2 \dfrac{\beta_{i\parallel}}{2}}}.
\end{equation}

The currents introduced above depend on the anisotropy and are thus affected by current sheet scattering. 
Using the relation between the curvature radius $R_c \simeq B_z/\mu_0 j_y$, and considering ions with $v_\bot = v_{iT\bot}$ so that $\kappa$ characterizes the bulk of the ion VDF, we have $\kappa^2 = R_c/\rho_i$, where $\rho_i = v_{iT\bot}/\omega_{ci}$. 
The latter can be written using Eq.~\ref{eq:j-tot} as
\begin{equation}
    \label{eq:kappa-tot}
    \kappa ^2  = \dfrac{1}{\beta_{i\parallel}} \dfrac{T_{i\parallel}}{T_{i\bot}}\dfrac{1 - \dfrac{n_c}{n_i}  (A_c - 1) \eta_c^2 \dfrac{\beta_{i\parallel}}{2}}{{\eta _e  + \dfrac{{n_s }}{n_i}\eta_s}}.
\end{equation}

In the limit of isotropic ions ($n_s/n_i = n_c/n_i \ll 1$, $T_{i\parallel} / T_{i\bot} = 1$), we obtain $j_y=\eta_e j_0$, and $\kappa^2 = 1 / \eta_e \beta_i = d_i^2 / \eta_e \rho_i^2$, i.e., $R_c = d_i^2 / \eta_e \rho_i$, where $d_i = \sqrt{m_i/ \mu_0 n_i e^2}$ is the ion inertial length. 
Below, we investigate the limits of parallel and perpendicular ion anisotropy, so that $j_y$ remains limited, ensuring stability of the current sheet.

\subsection{Limits for the ion anisotropy}
We consider two threshold regimes for the ion anisotropy $T_{i\bot}/T_{i\parallel}$. 
For $T_{i\perp}/T_{i\parallel} < 1$, the ion VDF is composed of a mixture of hot and cold ions. 
In contrast, for $T_{i\perp}/T_{i\parallel} > 1$, the ion VDF consists of hot and Speiser ions.

\subsubsection{Cold ions dominated regime}
In the first case, we consider that the ion population consists mainly of hot isotropic and cold ions, whereas Speiser ions are negligible ($n_s \approx 0$, $j_{sy} \approx 0$). 
Since cold ions are field-aligned counter-streaming beams, this regime corresponds to a parallel anisotropy with
\begin{equation}
    \left(\dfrac{T_{i\parallel}}{T_{i\bot}} - 1\right)  \approx \dfrac{n_c}{n_i} \left (A_c - 1 \right ) > 0,
\end{equation}
\noindent where we have assumed $T_{i\bot} \approx T_{c\bot}$. 
Thus, the current density Eq.~\ref{eq:j-tot} becomes
\begin{equation}
    \label{eq:j-tot-cold}
    j_y = j_0 \dfrac{\eta_e}{{1 - \left(\dfrac{T_{i\parallel}}{T_{i\bot}} - 1\right) \eta_c^2 \dfrac{\beta_{i\parallel}}{2}}}, 
\end{equation}
\noindent and the adiabaticity parameter
\begin{equation}
    \label{eq:kappa-cold}
    \kappa ^2  = 
    \dfrac{1}{\beta_{i\parallel}}
    \dfrac{T_{i\parallel}}{T_{i\bot}}\dfrac{1}{\eta _e} \left [ 1 - \left(\dfrac{T_{i\parallel}}{T_{i\bot}} - 1\right) \eta_c^2 \dfrac{\beta_{i\parallel}}{2}\right ].
\end{equation}
To prevent scattering of the regular adiabatic orbits of cold ions into chaotic orbits, which would otherwise reduce the current they carry, the magnetic field curvature radius $R_c$ must remain large compared with the ion gyroradius. 
In the opposite limit, the $\kappa = 0$ solution corresponds to scattering of cold ions into Speiser-type orbits, leading to an unlimited increase of the current density. 
This condition, therefore, defines the upper limit for the parallel anisotropy.
\begin{equation}
    \label{eq:threshold-cold}
    \mathcal{R}_c(\eta_c, \beta_{i\parallel}) = \left ( 1 + \dfrac{2}{{\eta_c^2\beta_{i\parallel}}}  \right )^{-1},
\end{equation}
\noindent where $\mathcal{R}_c(\eta_c, \beta_{i\parallel})$ denotes the anisotropy at the stability threshold. 
In the limit $\eta_c^2 = T_{c\bot}/T_{i\parallel} = 1$, we recover the classic fluid firehose stability condition where the ion inertia balances the magnetic tension force~\cite{cowley_plasma_1980}.

\begin{figure}[!t]
    \centering
    \includegraphics[width=\linewidth]{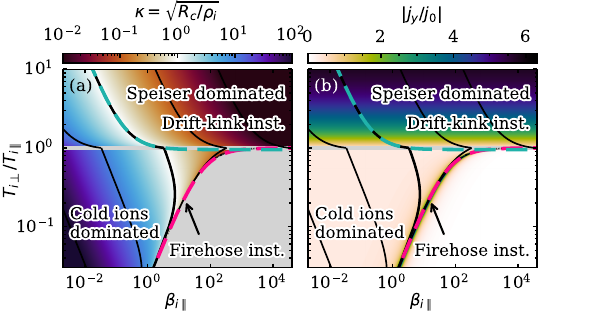}
    \caption{
        (a) Adiabaticity parameter and (b) current density in the model 1D current sheet with three ion populations with $\eta_e=0.3$, $\eta_s=0.2$, and $\eta_c=3$. 
        The parallel anisotropy limit [Eqs.~\ref{eq:j-tot-cold} and \ref{eq:kappa-cold}] is shown in the lower half of each panel $T_{i\bot}/T_{i\parallel} < 1$, and the perpendicular anisotropy limit [Eqs.~\ref{eq:j-tot-speiser} and \ref{eq:kappa-speiser}] is shown in the upper half $T_{i\bot}/T_{i\parallel} > 1$. 
        The black thick line indicate $\kappa=1$ and the black thin lines indicate $\kappa=10^{-2},\,10^{-1},\,10^{1},\,10^{2}$. 
        The dashed cyan and pink lines indicate the anisotropy thresholds [Eqs.~\ref{eq:threshold-cold} and~\ref{eq:threshold-speiser}].}
    \label{fig:model}
\end{figure}

We plot the adiabaticity parameter [Eq.~\ref{eq:kappa-cold}] and the current density [Eq.~\ref{eq:j-tot-cold}] for the cold ions dominated regime in the lower regions $T_{i\bot}/T_{i\parallel}< 1$ of Figure~\ref{fig:model}(a) and (b), respectively. 
We find that the adiabaticity parameter decreases with $\beta_{i\parallel}$. 
In particular, near the threshold [Eq.~\ref{eq:threshold-cold}] indicated by the lower dashed turquoise line, $\kappa$ drops abruptly, corresponding to the unlimited increase of current density $j_y$ when the denominator of Eq.~\ref{eq:j-tot-cold} becomes $0$.

\subsubsection{Speiser dominated regime}
In the second regime, we consider that cold ions provide a negligible current ($n_c \approx 0$, $j_{cy} \approx 0$), and Speiser ions carry the current. 
Hence, this regime corresponds to perpendicular temperature anisotropy with 
\begin{equation}
    \label{eq:a-speiser-dominated}
    \left(\dfrac{T_{i\parallel}}{T_{i\bot}} - 1\right)  \approx \dfrac{n_s}{n_i} \left (A_s - 1 \right ) < 0,
\end{equation}
\noindent where we have assumed $T_{i\bot} \approx T_{s\bot}$. 
Substituting Eq.~\ref{eq:a-speiser-dominated} into Eq.~\ref{eq:j-tot}, the current density becomes 
\begin{equation}
    \label{eq:j-tot-speiser}
    j_y = j_0 \left [\eta_e + 2 \left(1 - \dfrac{T_{i\parallel}}{T_{i\bot}}\right) \eta_s\right ],
\end{equation}
\noindent where we have used the approximation of the Speiser ion VDF~\cite{burkhart_ion_1992,burkhart_particle_1992,artemyev_ion_2019}, with $A_s \approx 1/2$. 
The corresponding adiabaticity parameter becomes
\begin{equation}
    \label{eq:kappa-speiser}
    \kappa ^2  = 
    \dfrac{1}{\beta_{i\parallel}}
    \dfrac{T_{i\parallel}}{T_{i\bot}}\dfrac{1}{\eta_e + 2 \left(1 - \dfrac{T_{i\parallel}}{T_{i\bot}}\right) \eta_s}.
\end{equation}

To avoid the scattering of hot ions into a quasi-adiabatic Speiser regime (when the entire ion population would move with the drift speed exceeding thermal speed and the current sheet would be strongly unstable), the magnetic field curvature parameter $\kappa$ must remain close to unity~\cite{buchner_regular_1989,chen_nonlinear_1992}. 
Thus, $\kappa = 1$ represents the marginal stability condition for the perpendicular anisotropy. 
Therefore, from Eq.~\ref{eq:kappa-speiser}, we obtain the second threshold
\begin{equation}
    \label{eq:threshold-speiser}
    \mathcal{R}_s(\eta_e, \eta_s, \beta_{i\parallel}) = \dfrac{1}{1 + \eta_e/2\eta_s} \left(1 + \dfrac{1}{2\eta_s} \dfrac{1}{\beta_{i\parallel}} \right).
\end{equation}
In the limit $\eta_s \ll 1$ ($v_{sD} \ll v_{iT\bot}$), the Speiser population is embedded within the hot ion population, and the threshold reduces to $\mathcal{R}_s(\eta_e, \eta_s \rightarrow 0, \beta_{i\parallel}) \rightarrow 1 / \eta_e \beta_{i\parallel}$. 
In this regime, magnetic field stretching is governed by the electron current such that for a given $\beta_{i\parallel}$ and $T_{i\bot}/T_{i\parallel}$, ions with $v_{iT\bot} > V_{Ai} / \sqrt{\eta_e}$ will scatter into a Speiser-type orbit. 
In contrast, for $\eta_s \gg 1$ ($v_{sD} \gg v_{iT\bot}$), the threshold approaches $\mathcal{R}_s(\eta_e, \eta_s \rightarrow \infty, \beta_{i\parallel}) \rightarrow 1$, corresponding to the most unstable configuration with a strong drift current carried by quasi-adiabatic Speiser orbits, characteristic of a drift-type instability~\cite{lapenta_kinetic_1997,daughton_unstable_1999}.

\section{Spacecraft Observations}
\label{sec:spacecraft-observations}
We validate our model using \textit{in situ} spacecraft observations in the Earth's magnetotail. 
We employ measurements in the near-Earth magnetotail ($X_{GSM}> -29~R_E$, in Geocentric Solar Magnetospheric (GSM) coordinates) from the Magnetospheric Multiscale (MMS) mission~\cite{burch_magnetospheric_2016} and in the mid-tail at lunar distances ($X_{GSM} \sim - 60~R_E$) from the Acceleration, Reconnection, Turbulence, and Electrodynamics of the Moon's Interaction with the Sun (ARTEMIS) mission~\cite{angelopoulos_artemis_2011}.

\subsection{Near-Earth reconnection}
The MMS dataset includes 516 fast plasma flows measured between 2017 and 2021, at $X_{GSM} \geq -29~R_E$, in the central plasma sheet corresponding to $\beta_i \geq 0.5$, where $\beta_i = 2 \mu_0 k_B T_i / B^2$ is the ion plasma beta~\cite{richard_are_2022}. 
Magnetic field measurements are obtained from the fluxgate magnetometer~\cite{russell_magnetospheric_2016}, while ion and electron VDFs and their moments are provided by the Fast Plasma Investigation instrument~\cite{pollock_fast_2016}. 
We correct the ion VDFs for penetrating radiation using the empirical background model described in Ref.~\onlinecite{gershman_systematic_2019}. 
To improve counting statistics and reduce uncertainties in moment calculations~\cite{gershman_calculation_2015}, we average three consecutive ion VDFs over a 450 ms window. 
It yields a total of 235,635 ion VDFs for analysis.

\begin{figure}[!t]
    \centering
    \includegraphics[width=\linewidth]{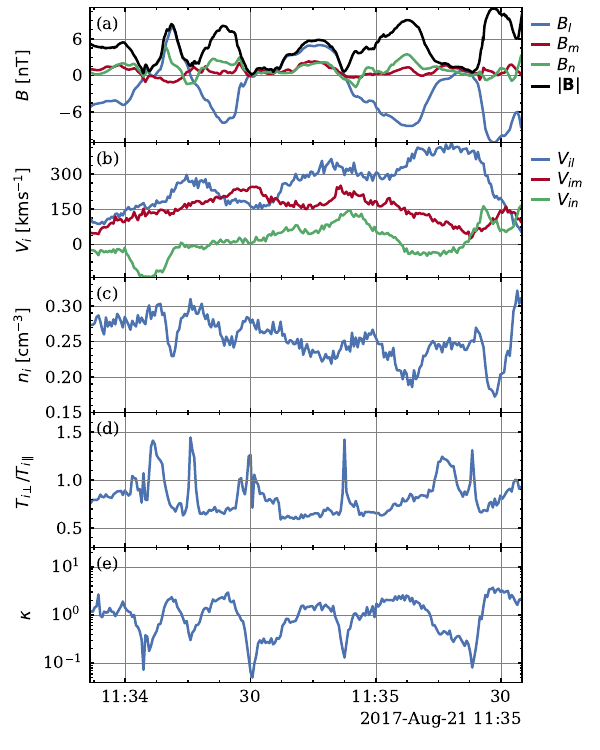}
    \caption{
        Overview of the example event from MMS. 
        (a) Magnetic field in the minimum variance. 
        (b) Ion bulk velocity in GSM coordinates. 
        (c) Ion number density. 
        (d) Ion temperature anisotropy with respect to the local magnetic field. 
        (e) Curvature parameter $\kappa$.}
    \label{fig:event}
\end{figure}

\begin{figure}[!b]
    \centering
    \includegraphics[width=\linewidth]{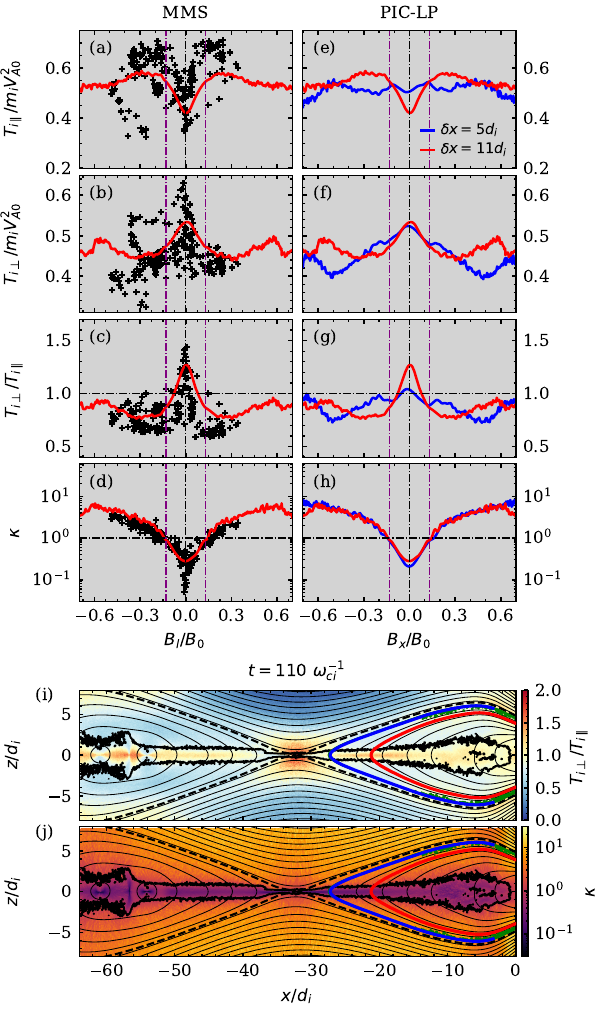}
    \caption{
        Comparison between MMS observations and PIC simulation results for ion temperature anisotropy. 
        Panels (a)–(d) and (e)–(h) show 1D profiles of: (a, e) ion temperature parallel to the local magnetic field, (b, f) ion temperature perpendicular to the magnetic field, (c, g) ion temperature anisotropy, and (d, h) the magnetic field curvature parameter. 
        Black crosses represent MMS measurements during the event shown in Fig.~\ref{fig:event}. 
        Solid blue and red lines correspond to profiles extracted along magnetic field lines in the PIC simulation, intersecting the $z=0$ plane at $\delta x = 5d_i$ and $\delta x = 11d_i$, respectively.
        Panels (i) and (j) present 2D maps of ion temperature anisotropy and curvature parameter from the PIC simulation. 
        The black contour indicates $\kappa=1$. 
        The black, thick-dashed lines indicate the separatrices. 
        The green contour indicates the region of interest for Fig.~\ref{fig:brazil-plots}(c).}
    \label{fig:mms-pic}
\end{figure}

To illustrate the limiting cases of cold ion-dominated regime, corresponding to parallel anisotropy, and Speiser ion-dominated, corresponding to perpendicular anisotropy, we examine a reconnection outflow observed by MMS on 21 August 2017\cite{richard_fast_2023} [Fig.~\ref{fig:event}]. 
During this interval, the magnetic field in Fig.~\ref{fig:event}(a) displays large-amplitude oscillations attributed to a flapping motion of the magnetotail current sheet~\cite{richard_observations_2021}, so that MMS samples across the reconnection outflow shown in Fig.~\ref{fig:event}(b). 
Here, we rotated the magnetic field and the ion bulk velocity to the $\bm{\hat{l}}\bm{\hat{m}}\bm{\hat{n}}$ coordinates system, where $\bm{\hat{l}}= \left [0.97,\,-0.25\,0.02 \right ]\,\textrm{GSM}$ is the maximum variance direction obtained from maximum variance analysis (MVA) of the magnetic field with $\lambda_{max}/\lambda_{int} = 15$, $\bm{\hat{n}}= (\bm{\hat{l}}\times \bm{j}) / |(\bm{\hat{l}}\times \bm{j})|$, and $\bm{\hat{m}} = \bm{\hat{n}}\times \bm{\hat{l}}$~\cite{sergeev_survey_2006}. 
In this coordinate system, at $|B_l| \approx 0$, MMS is around the center of the current sheet, while maxima in $|B_l|$ correspond to the current sheet edges. 
We present the ion temperature anisotropy $T_{i\bot}/T_{i\parallel}$ [Fig.~\ref{fig:event}(d)] with respect to the local magnetic field, where we calculated $T_{i\bot}$ and $T_{i\parallel}$ rotating the temperature tensor $\mathbf{T}_i$ in the magnetic field-aligned (MFA) frame~\cite{swisdak_quantifying_2016}. 
In this frame, one axis aligns with the local magnetic field direction $\hat{\bm{b}}$. 
The perpendicular directions are set to equalize the remaining diagonal components of the ion temperature tensor $\mathbf{T}_i = T_{i\parallel} \hat{\bm{b}}\hat{\bm{b}} + T_{i\bot} (\mathbf{1} - \hat{\bm{b}}\hat{\bm{b}}) + \bm{\Theta}_i$, where $T_{i\parallel}$ and $T_{i\bot}$ are the parallel and perpendicular temperatures, and $\bm{\Theta}_i$ is the agyrotropic part with zero diagonal. 
We compute the magnetic field curvature radius $R_c$, and the corresponding adiabaticity parameter $\kappa$, using the multi-spacecraft curlometer technique~\cite{chanteur_spatial_1998,chanteur_spatial_1998a}. 
We find that the temperature anisotropy [Fig.~\ref{fig:event}(d)] and the adiabaticity parameter [Fig.~\ref{fig:event}(e)] vary strongly across the reconnection outflow and are inversely correlated, indicating interplay between magnetic field curvature and the ion VDFs. 
To examine the spatial variation of the temperature anisotropy across the current sheet, we construct profiles of $T_{i\parallel}$ and $T_{i\bot}$ as functions of the reconnecting magnetic field component $B_l$ [Figs.~\ref{fig:mms-pic}(a)-~\ref{fig:mms-pic}(d)]. We normalize $B_l$ by the background (lobe) magnetic field strength $B_0 = B \sqrt{1 + \beta_i}$, estimated from pressure balance of the magnetotail plasmasheet~\cite{asano_evolution_2003}. 
To facilitate comparison with numerical simulation (see Section~\ref{sec:numerical-sim}), we normalize the ion temperatures $T_i$, $T_{i\parallel}$, $T_{i\bot}$ to the electromagnetic energy available per particle $ m_iV_{A0}^2$, where $V_{A0} = B_0 / \sqrt{\mu_0 m_i n_0}$ is the Alfv\'en speed and $n_0$ the ion density within the current sheet. 
For the flapping event presented in Fig.~\ref{fig:event}, we obtain $B_0 = 21.6\,\mathrm{nT}$ and $n_0 = 0.27\,\mathrm{cm}^{-3}$, the latter computed as the average number density over intervals with $|B_l| / B_0 < 0.1$ well within the current sheet.
At the edges of the current sheet -- corresponding to where $B_l / B_0 = \pm 0.3$ -- the ion VDFs are elongated parallel to the magnetic field, with $T_{i\bot}/T_{i\parallel} < 1$, characteristic of cold, field-aligned counter streaming beams~\cite{hoshino_ion_1998,hietala_ion_2015} as described in our model. 
We find that instances of parallel anisotropy coincide with $\kappa > 1$, indicating that the cold ions follow regular adiabatic orbits. 
These observations suggest that outside the current sheet center, the ion VDF primarily consists of an adiabatic, parallel-anisotropic population, as described by our model.

In contrast, around the current sheet center, where the magnetic field reverses, the ion VDFs exhibit a perpendicular temperature anisotropy, $T_{i\bot}/T_{i\parallel} > 1$, consistent with Speiser-type ion VDFs~\cite{nakamura_ion_1998,lottermoser_ion_1998,hietala_ion_2015}. 
Co-located with this perpendicular temperature anisotropy, the curvature parameter is $\kappa \ll 1$, indicating that ions are following quasi-adiabatic trajectories, such as in Speiser orbits. 
Using Eq.~\ref{eq:a-speiser-dominated} with the observed $T_{i\bot}/T_{i\parallel} \approx 1.4$, we obtain $n_s/n_i\approx 0.57$ within the upper range of values reported in previous observations~\cite{artemyev_proton_2010}. 
Hence, the MMS observations indicate that within the current sheet, the ion current is carried by a quasi-adiabatic, perpendicular, anisotropic population, consistent with the Speiser population used in the model. 

\begin{figure}[!t]
    \centering
    \includegraphics[width=\linewidth]{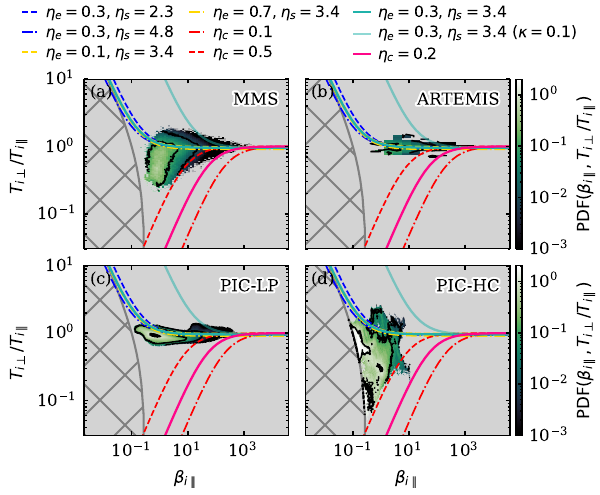}
    \caption{
        Distribution $\textrm{PDF}(T_{i\bot}/T_{i\parallel}, \beta_{i\parallel})$ in: (a) the near-Earth magnetotail BBFs from MMS~\cite{richard_fast_2023}, (b) the midtail current sheet~\cite{kamaletdinov_thin_2024}, (c) PIC-LP simulation of forced magnetic reconnection in a Lembege and Pellat equilibrium magnetotail current sheet~\cite{an_configuration_2022}, and (d) PIC-HC simulation of forced reconnection in a Harris-like equilibrium current sheet with cold ions~\cite{norgren_presence_2021}. 
        The dark blue and yellow dashed and dash–dotted lines show the thresholds given by Eq.~\ref{eq:threshold-speiser} for the extreme values of the confidence intervals of $\eta_s$ and $\eta_e$, respectively. 
        The red dashed and dash–dotted lines indicate the thresholds from Eq.~\ref{eq:threshold-cold} for the extreme values of the confidence interval of $\eta_c$. 
        Solid cyan and pink lines mark the thresholds corresponding to the nominal values $(\eta_e = 0.4, ~\eta_s = 3)$, and $\eta_c = 0.2$, respectively. 
        The translucent cyan line denotes the threshold for perpendicular temperature anisotropy modified for $\kappa = 0.1$.}
    \label{fig:brazil-plots}
\end{figure}

To assess whether the theoretical limits (Eqs.~\ref{eq:threshold-cold} and~\ref{eq:threshold-speiser}) bound the observed anisotropy, we present the joint probability density function (PDF) of $\beta_{i\parallel}$ and $T_{i\bot}/T_{i\parallel}$, for the dataset of 235,635 ion VDFs, in Fig.~\ref{fig:brazil-plots}a. 
It was previously suggested that, for this dataset, curvature scattering is responsible for ion isotropization~\cite{richard_fast_2023}. 
We estimate the electron current fraction $\eta_e = j_{ey} / j_{0}$, with $j_{ey}$ the electron current calculated using the measured number density and electron bulk velocity, and $j_{0}$ calculated using the measured number density and ion temperature. 
We find a median value of $\eta_e = 0.3$ with a $95\%$ confidence interval $\eta_e \in [0.1, 0.7]$. 
To estimate the ratio of drift speed of Speiser ions to thermal speed, we compute $\eta_s =  |\nabla \times \bm{B}| / \mu_0 j_0 (n_s/n_i)$, at the center of the current sheet ($B/B_0 < 0.1$), where Speiser ions dominate the current contribution. 
We compute $|\nabla \times \bm{B}|$ using the curlometer technique~\cite{chanteur_spatial_1998,chanteur_spatial_1998a}, and adopt a typical fraction of Speiser ions $n_s/n_i \simeq 0.15$~\cite{artemyev_proton_2010}. 
It yields a median value of $\eta_s = 3.4$ with a $95\%$ confidence interval of $\eta_s \in [2.3, 4.8]$. 
Estimating the cold ion current fractions, $\eta_c$, from the full dataset is challenging due to difficulties in isolating the cold ion population in the ion VDF and in accurately measuring its temperature~\cite{li_quantification_2021}. 
Instead, we adopt representative values of the cold ions perpendicular temperature $T_{c\bot}$ consistent with prior studies, choosing $T_{c\bot}\in [200~\textrm{eV}, \, 600~\textrm{eV}]$~\cite{li_quantification_2021}, and $T_{i\parallel}\in[1~\textrm{keV}, \, 6~\textrm{keV}]$ measured in the dataset at the current sheet edges $|B|/B_0 \in [0.2, 0.7]$, so that $\eta_c\in [0.1, 0.5]$. 
We find that the observed distribution of temperature anisotropy is consistent with the predicted limits for current-sheet stability, particularly in the parallel-anisotropy regime. 
Although a substantial fraction of the dataset exhibits perpendicular anisotropies that exceed the threshold given by Eq.~\ref{eq:threshold-speiser}, these cases rarely surpass the limit corresponding to $\kappa = 0.1$ (light cyan line), obtained by replacing $\kappa^2 = 1$ with $\kappa^2 = 0.01$ in the derivation of Eq.~\ref{eq:threshold-speiser}. 
Even for such values of $\kappa \in [0.1, 1]$, ions undergo strong diffusion $\Delta \mathcal{I}_z / \mathcal{I}_z \sim \kappa$~\cite{buchner_regular_1989}. 
Hence, in analogy to the thresholds for temperature anisotropy-driven instabilities, the $\kappa=1$ corresponds to the marginal stability, while the $\kappa = 0.1$ threshold corresponds to a faster growth rate. 
Therefore, the MMS observations support well the thresholds derived in Eqs.~\ref{eq:threshold-cold} and~\ref{eq:threshold-speiser}.

\subsection{Lunar distances}
To complement MMS observations, we use data from the Acceleration, Reconnection, Turbulence, and Electrodynamics of the Moon's Interaction with the Sun (ARTEMIS) mission. 
ARTEMIS consists of two spacecraft orbiting the Moon, enabling measurements of the magnetotail at lunar distances ($X_{GSM} \sim -60R_{E}$). 
The magnetotail at such distances often exhibits mixed plasma compositions, with both hot and suprathermal ($\gtrsim 1$ keV) ions originating from near-Earth and distant-tail reconnection, as well as cold ($<100$ eV) ions of ionospheric or mantle origin~\cite{wang_properties_2014,runov_properties_2023}. 
At these distances, the dipole contribution to the equatorial magnetic field is essentially nonexistent, allowing for a significant fraction of demagnetized (hot) ions executing Speiser-like motion to coexist with magnetized cold ions undergoing $\mathbf{E}\times\mathbf{B}$ motion~\cite{kamaletdinov_ion_2025}. 
Thus, the ARTEMIS dataset offers a unique opportunity to study both the effects of Speiser-like ions and the Hall electric field that governs cold-ion dynamics. 
The resultant dataset includes 1261 magnetotail current sheet crossings collected between 2012 and 2022~\cite{kamaletdinov_thin_2024}, spanning $X_{GSM} \in [-65, -50]R_E$. 
Magnetic field measurements are provided by the fluxgate magnetometer~\cite{auster_themis_2008}, and ion parallel and perpendicular temperatures are measured by electrostatic analyzers~\cite{mcfadden_themis_2008}.

We plot the joint PDF of $\beta_{i\parallel}$ and $T_{i\bot}/T_{i\parallel}$ in Fig.~\ref{fig:brazil-plots}(b). 
We find that the data mostly fall at higher-$\beta_{i\parallel}$ compared with near-Earth magnetotail [Fig.~\ref{fig:brazil-plots}(a)], due to the weaker magnetic field and larger temperature, and nearly isotropic ion VDFs $T_{i\bot}/T_{i\parallel} \approx 1$. 
Although the data are very concentrated in the high $\beta_{i\parallel}$ and $T_{i\bot}/T_{i\parallel} \approx 1$ end, the distribution resembles that of MMS data [Fig.~\ref{fig:brazil-plots}a], and the iso-contours also qualitatively resemble the thresholds, providing further support for these limits.

\section{Numerical simulations}
\label{sec:numerical-sim}
To further validate the limits for the ion anisotropy obtained in Eqs.~\ref{eq:threshold-cold} and~\ref{eq:threshold-speiser}, we analyze two fully kinetic particle-in-cell (PIC) simulations of symmetric magnetic reconnection in a magnetotail-like configuration~\cite{an_configuration_2022,norgren_presence_2021}.

\subsection{Lemb\`ege–Pellat equilibrium}
The first simulation from Ref.~\onlinecite{an_configuration_2022}, hereafter PIC-LP, employs a generalized Lemb\`ege–Pellat current sheet equilibrium~\cite{lembege_stability_1982,lu_hall_2019}, which includes a background plasma population and a polarization electric field, typical of magnetotail current sheets~\cite{artemyev_thin_2009,runov_local_2006}.

\subsubsection{Numerical setup}
We perform a PIC simulation~\cite{pritchett_kinetic_2001,pritchett_externally_2005} of driven magnetic reconnection starting from the Lemb\`ege–Pellat equilibrium described below in Section~\ref{sssec:initial-equilibrium-lp}. 
The simulation is two-dimensional in configuration space (2D) and three-dimensional in velocity space (3V). 
Normalizations are based on ion quantities: time is normalized to the inverse ion cyclotron frequency $\omega_{ci} = e B_0 / m_i$, lengths to the ion inertial length $d_i = \sqrt{m_i/ \mu_0 n_0 e^2}$, velocities to the Alfv\'en speed $V_{A0}$, and energies to $m_i V_{A0}^2$. 
The simulation domain extends over $[-L_z/2, L_z/2] \times [-L_x, 0]$, with $L_x=64 \, d_i$ and $L_z=16\, d_i$, and a grid cell size of $\delta/d_i = 1 / 32$. 
The Debye length is $\lambda_{De} = 0.46\, \delta$, the ion-to-electron mass ratio is $m_i / m_e = 100$, and the reduced speed of light is $c / V_{A0} = 20$. 
The time step is $\Delta t = 0.001 \, \omega_{ci}^{-1}$. 
The reference density $n_0$ is represented by 425 particles per cell, corresponding to a total of $\sim 2 \times 10^8$ particles initially. 
The simulation domain has open boundary conditions at $x/d_i=-L_x$ and $x/d_i=0$ that allow particles to exit the domain while continuously injecting back particles in the system~\cite{aldrich_particle_1985,pritchett_externally_2005}, and reflective boundary conditions at $z/d_i=-L_z/2$ and $z/d_i=L_z/2$.

\subsubsection{Initial equilibrium}
\label{sssec:initial-equilibrium-lp}
The initial equilibrium is obtained by self-consistently solving for the vector potential $A_y(\varepsilon x, z)$ and the electrostatic potential $\phi(\varepsilon x, z)$, where $|\varepsilon|\ll 1$ indicates weak inhomogeneity in the $x$ direction. 
Ampere's Law and the quasi-neutrality condition, respectively, read:
\begin{equation}
    \label{eq:ampere-lp}
    \dfrac{\partial^2 A_y}{\partial z^2} = -\mu_0 \sum_{\alpha} q_{\alpha} n_0 v_{\alpha D} \exp \left (-\dfrac{q_{\alpha}\phi}{T_{\alpha 0}} + \dfrac{q_{\alpha}v_{\alpha D}A_y}{T_{\alpha 0}}\right ),
\end{equation}
\begin{multline}
    \label{eq:quasi-neutrality-lp}
    \sum_{\alpha} q_{\alpha} n_0 \exp \left (-\dfrac{q_{\alpha}\phi}{T_{\alpha 0}} + \dfrac{q_{\alpha}v_{\alpha D}A_y}{T_{\alpha 0}}\right )\\ + \sum_{\alpha} q_{\alpha} n_b \exp \left (-\dfrac{q_{\alpha}\phi}{T_{\alpha b}}\right ) = 0,
\end{multline}
\noindent where we have separated the background population from the current-carrying population in the current sheet. 
Here, $\alpha=i,e$ denotes the particle species, $q_\alpha$ the particle charge, $n_0$ and $n_b$ the number densities of the current sheet and background populations, respectively, $T_{\alpha 0}$ and $T_{\alpha b}$ the current sheet and background species temperatures, and $v_{\alpha D}$ the drift speed. 
We use $n_b/n_0=1/5$, $T_{i0}=T_{ib}=5m_iV_{A0}^2/12$, $T_{e0}=T_{eb}=m_iV_{A0}^2/12=T_{i0}/5$, $v_{iD}=V_{A0}/3$ and $v_{eD}=-5V_{A0}/3$. 
Solving Eqs.~\ref{eq:ampere-lp} and~\ref{eq:quasi-neutrality-lp}, with the number densities and current densities calculated using a drifting Maxwellian equilibrium VDF~\cite{lembege_stability_1982}, and the boundary conditions at the current sheet center $z=0$ where $B_x=0$ and $B_z = \varepsilon B_0$:
\begin{equation}
    \left . \frac{\partial A_y}{\partial z} \right |_{z=0} = 0, \qquad A_y(z=0) = \varepsilon B_0 x,
\end{equation}
\noindent we obtain the initial equilibrium for the simulation.

This initial configuration, featuring a finite magnetic field component normal to the current sheet ($B_z \neq 0$), is stable against spontaneous reconnection~\cite{pellat_does_1991}. 
We initiate reconnection by imposing electric fields:
\begin{equation}
    \label{eq:driving-ey}
    E_{y,drive}(x, z=\pm L_z/2, t) = E_{y0}f(t)\sin^2 \left ( \frac{\pi x}{L_x}\right ),
\end{equation}
\noindent at the $z/d_i=\pm L_z/2$ boundaries. 
Here, $E_{y0}/E_0 = 0.4$, and the $f(t)=\tanh \left ( \omega t\right )  \operatorname{sech}^2 \left ( \omega t\right )$ with the characteristic frequency $\omega / \omega_{ci} = 20 $. 
This forcing drives plasma inflows that compress and thin the current sheet, therefore forcing magnetic reconnection.

\subsubsection{Results}
Figure~\ref{fig:mms-pic}(i) shows the ion anisotropy $T_{i\bot}/T_{i\parallel}$ at $t\omega_{ci} = 110$, slightly after the peak of the reconnection rate, which occurs at $t\omega_{ci}=84$ (see Figure 1 in Ref.~\onlinecite{an_configuration_2022}). 
Here, we calculate $T_{i\bot}$ and $T_{i\parallel}$ using the rotation to the MFA as described in Section~\ref{sec:spacecraft-observations}. 
We note that the extreme parallel anisotropies at the top and bottom of the domain arise from the reflective boundary conditions and are not physical. 
In the reconnection outflows, the parallel temperature dominates outside the current sheet, whereas the perpendicular temperature dominates at the current sheet center. 
In particular, we compute the adiabaticity parameter $\kappa$ [Fig.~\ref{fig:mms-pic}(j)], and overlay the contour of $\kappa=1$ in Fig.~\ref{fig:mms-pic}(i). 
We find that the contour of $\kappa=1$ clearly separates the parallel anisotropy outside the current sheet, where $\kappa > 1$, and the perpendicular anisotropy inside the current sheet, where $\kappa \lesssim 1$. 
This is consistent with the MMS observations [Fig.~\ref{fig:event}], and further supports our model.

To further compare the PIC-LP simulation with MMS measurements across the current sheet [Figs.~\ref{fig:event}(a)-~\ref{fig:event}(b)], we extract the parallel and perpendicular temperatures along two field lines crossing the neutral plane at $\delta x = 5\, d_i$ and $\delta x = 10\, d_i$, where $\delta x$ denotes the distance downstream from the X-line at the center of the current sheet. 
We find that the field line at $\delta x = 10\, d_i$ closely reproduces the observed trend, indicating that the PIC-LP simulation accurately captures the magnetotail dynamics. 
In particular, we observe that near $B_x/B_0=0$, the ion perpendicular temperature dominates ($T_{i\bot}/T_{i\parallel} > 1$), as expected for Speiser ions. 
In contrast, at the current sheet edges, the parallel temperature dominates ($T_{i\bot}/T_{i\parallel} < 1$), as expected for field-aligned beams. 
The corresponding adiabaticity parameter $\kappa$ shows excellent qualitative and quantitative agreement with the MMS data. 
We note that, at the center of the current sheet $B_x/B_0\approx 0$, the observed $\kappa$ reaches smaller values than those found in the PIC-LP simulation, which may be attributed to finite grid resolution and the artificial mass ratio leading to an artificially thicker electron-scale embedded current sheet ($d_e = \sqrt{m_e/m_i}d_i$). 

We present the joint PDF of $\beta_{i\parallel}$ and $T_{i\bot}/T_{i\parallel}$ in the PIC-LP simulation in Fig.~\ref{fig:brazil-plots}(c). 
To compare the PIC-LP simulation with MMS observations, we restrict to regions in the outflow where $\beta_i>0.1$ and $V_{ix} > 0$ (green contour in Figs.~\ref{fig:mms-pic}(i) and~\ref{fig:mms-pic}(j)). 
To collect a statistically significant number of points, we combine data within $t\omega \in [84, 128]$ during which the reconnection electric field $E_{y,rec}$ at the X-line monotonically decreases from its peak value $E_{y,rec}^{max}/E_0=0.07$ at $t\omega_{ci}=84$ but remains $E_{y,rec} / E_{y,rec}^{max} > 1/2$. 
We find that the distribution of ion anisotropy in the PIC-LP simulation is relatively similar to that observed by MMS. 
However, we find only weak parallel anisotropy, possibly due to the initially hot background with $T_{ib} = T_{i0}$ and the absence of initially cold ions. 
Nevertheless, we find that the threshold $\mathcal{R}_s$ bounds the distribution well, supporting the derived thresholds.

\subsection{Harris equilibrium with cold ions}
The second simulation from Ref.~\onlinecite{norgren_presence_2021}, hereafter referred to as PIC-HC, employs a Harris current sheet equilibrium with cold ions as the inflow populations. 
While this configuration is slightly more idealized and suited to study anti-parallel magnetic reconnection, the inclusion of initially cold ions provides the opportunity to test the limit for parallel anisotropy.

\subsubsection{Numerical setup}
We perform a particle-in-cell (PIC) simulation~\cite{hesse_diffusion_1999} of driven magnetic reconnection starting from the Harris equilibrium described in Section~\ref{sssec:initial-equilibrium-hc}. 
The simulation is 2D in configuration space and 3V in velocity space. 
Normalizations are those of the simulation of the PIC-LP run. 
The simulation domain extends over $[-L_z/2, L_z/2] \times [0, L_x]$, with $L_x=200 \, d_i$ and $L_z=12.5\, d_i$, divided into a grid of $6400 \times  1600$ cells. 
The ion-to-electron mass ratio is $m_i / m_e = 100$, and the reduced speed of light is $c / V_{A0} = 20$. 
The time step is $\Delta t = 0.5 \omega_{pe}^{-1}=0.0025 \omega_{ci}^{-1}$. 
The inflow density $n_c$ is represented by $\sim 30$ particles per cell, and the total number of particles is $\sim 4.8 \times 10^8$. 
The simulation domain has periodic boundary conditions at $x/d_i=0$ and $x/d_i=L_x$, and reflective boundary conditions at $z/d_i=-L_z/2$ and $z/d_i=L_z/2$.

\subsubsection{Initial equilibrium}
\label{sssec:initial-equilibrium-hc}
The initial equilibrium is a Harris equilibrium with two separate ion and electron populations: the preexisting current sheet (Harris) population and a cold inflow. 
The magnetic field in the Harris equilibrium is given by
\begin{equation}
    \label{eq:bx-harris}
    \bm{B} = B_0 \tanh \left (\frac{z}{L} \right ) \bm{\hat{x}},
\end{equation}
\noindent with $L$ the current sheet thickness. 
The corresponding equilibrium number density is 
\begin{equation}
    \label{eq:number-density-harris}
    n = n_0 \operatorname{sech}^2 \left (\frac{z}{L} \right ) + n_c \left [ \frac{1}{2} + \frac{1}{2} \tanh \left ( \frac{|z| - 2 L}{L / 2} \right ) \right ],
\end{equation}
\noindent where $n_0$ and $n_c$ are the number densities of the current sheet and the cold ions in the inflow, respectively. 
Here we use $n_c/n_0=1/5$, $L=2d_i$, $T_{i0}/T_{e0}=5$, and $T_{ic}=T_{ec}=0$. 
We initiate magnetic reconnection by forcing a localized perturbation at the center of the simulation domain.

\subsubsection{Results}
\begin{figure}[!t]
    \centering
    \includegraphics[width=\linewidth]{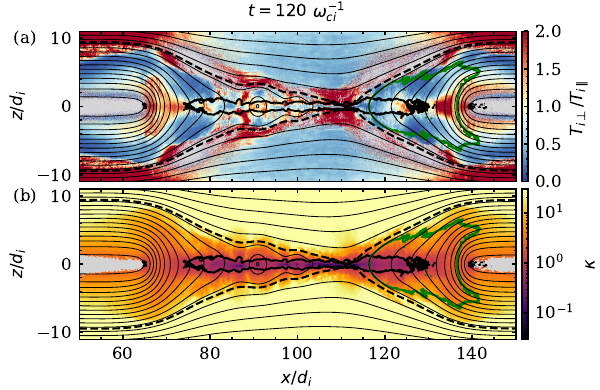}
    \caption{
        Anisotropy $T_{i\bot}/T_{i\parallel}$ (a) and adiabaticity parameter $\kappa$ (b) of initially cold ions in the PIC-HC simulation. 
        The black contour indicates $\kappa=1$. The black, thick-dashed lines indicate the separatrices. 
        The green contour indicates the region of interest for Fig.~\ref{fig:brazil-plots}(d).}
    \label{fig:pic-harris-cold}
\end{figure}
We present the anisotropy $T_{i\bot}/T_{i\parallel}$ of the initially cold ions at time $t=120~\omega_{ci}^{-1}$ in Fig.~\ref{fig:pic-harris-cold}. 
At the time and place of investigation, the effect of the periodicity of the simulation is negligible, and the reconnection exhaust is completely dominated by the initially cold ions~\cite{norgren_presence_2021}. 
Similar to that seen in Fig.~\ref{fig:mms-pic}(j), the temperature anisotropy is mainly parallel outside the current sheet, except in the inflow near the X-line, near the dipolarization front, and at ion mirror points. 
We compute the adiabaticity parameter $\kappa$ [Fig.~\ref{fig:pic-harris-cold}(b)], and overlay the contour of $\kappa=1$ in Fig.~\ref{fig:pic-harris-cold}(a). 
We find that, similar to that in the PIC-LP run [Fig.~\ref{fig:mms-pic}(i)], the anisotropy tends to be larger in regions where $\kappa \lesssim 1$ than in the vicinity outside of the current sheet where $\kappa > 1$.

We present the joint PDF of $\beta_{i\parallel}$ and $T_{i\bot}/T_{i\parallel}$ in Fig.~\ref{fig:brazil-plots}(d). 
Similar to the PIC-LP run and MMS observations, we restrict to regions of the outflow where $\beta_i>0.1$ and $V_{ix} > 0$ (green contour in Fig.~\ref{fig:pic-harris-cold}). 
Compared to the PIC-LP simulation, $T_{i\perp}/T_{i\parallel}$ spans a broader range. 
This is due to the inclusion of cold ions in the reconnection inflow, which undergo strong (larger than their initial temperature) acceleration in the reconnection outflow, leading to a pronounced parallel temperature anisotropy. 
In addition, the ion plasma beta remains small $\beta_{i\parallel}\lesssim 10$ in this simulation, allowing a broader range of temperature anisotropies than in higher-$\beta_{i\parallel}$ configurations, such as those inferred from ARTEMIS observations. 
Nevertheless, similar to MMS observations, the instability thresholds constrain the distribution relatively well for $T_{i\perp}/T_{i\parallel} < 1$. 
In contrast, the perpendicular anisotropy slightly exceeds the marginal stability condition $\kappa=1$. 
However, similar to MMS observations, this may be attributed to overestimates of $\eta_e$ and $\eta_s$, or weak scattering, i.e., slow growth of the instability.

\section{Discussion}
We derived two limits for the ion anisotropy in a magnetotail-like quasi-1D current sheet with a finite normal magnetic field, incorporating three ion populations: an adiabatic cold population of two counter-streaming beams, an isotropic hot population, and a quasi-adiabatic Speiser population. 
Here, we focused on current sheets in the reconnection outflow. 
Nevertheless, the assumptions of the model with a thin current sheet geometry, finite $B_z$, and the superposition of three ion populations are also commonly observed in quiet thin current sheets in the magnetotail~\cite{artemyev_ion_2019}, suggesting that the derived anisotropy limits are applicable in these configurations as well. 
The resulting limits define a maximum ion temperature anisotropy as a function of $\beta_{i\parallel}$, similar to the empirical thresholds associated with mirror, firehose, and proton cyclotron instabilities~\cite{yoon_kinetic_2017}. 
Unlike these instability criteria, however, the present limits arise from nonlinear particle dynamics within the current sheet, and may constrain the accessible anisotropy prior to the development of kinetic instabilities~\cite{richard_fast_2023}. 
Here, we discuss the physical interpretation of these limits and their implications for current sheet stability.

The first limit [Eq.~\ref{eq:threshold-cold}], which concerns the parallel anisotropy, resembles the firehose condition, in which the plasma pressure force balances the magnetic tension. 
In the particular case where $\eta_c = 1$, i.e., $T_{c\bot} = T_{i\parallel}$ so that the ion VDF consists of a single population with anisotropy $T_{i\bot}/T_{i\parallel}$, the threshold reduces to the well-known fluid firehose condition~\cite{cowley_plasma_1980}. 
In contrast, in the limit $\eta_c^2 \beta_{i\parallel} \rightarrow 0$, the threshold becomes $\mathcal{R}_c(\eta_c \rightarrow 0, \beta_{i\parallel}) \rightarrow \eta_c^2 \beta_{i\parallel}$. 
This regime corresponds to strongly magnetized cold ions, for which $T_{c\bot}/B^2 \ll 1$, implying that larger magnetic field curvature is required to scatter the cold ions. 
Consequently, cold ions can maintain force balance even in highly stretched magnetic field configurations.

The second limit [Eq.~\ref{eq:threshold-speiser}], which concerns the perpendicular anisotropy, defines the threshold above which the hot isotropic ion population is scattered into quasi-adiabatic transient Speiser orbits. 
Since the adiabaticity parameter for one particle inversely scales with the particle's perpendicular velocity $\kappa \propto 1 / \sqrt{v_\bot}$, for a given increase of the magnetic curvature due to e.g., current sheet thinning/stretching, the supra-thermal ions with the corresponding perpendicular velocity, initially in the hot isotropic background, will be scattered into Speiser orbits. 
As a result, the number density $n_h$ and temperature $T_h$ of the hot isotropic background will decrease, and concurrently, the number density $n_s$ of the quasi-adiabatic Speiser population will increase, thereby increasing their current density. 
Eventually, the resulting ion VDF will be composed of a core plus a dense drifting beam Speiser population. 
If the drift velocity is sufficiently large compared with the thermal speed and the drifting beam is sufficiently dense with respect to the core, this will lead to the growth of a drift-kink type instability~\cite{lapenta_kinetic_1997,daughton_unstable_1999,zelenyi_low_2009}. 

The drift-kink instability may act back on the ion VDFs, scattering quasi-adiabatic Speiser ions into the hot isotropic population. 
Recent \textit{in situ} observations suggest that the growth of the drift-kink instability may ultimately lead to a thickening of the current sheet~\cite{richard_observations_2021}. 
During the early stage of the drift-kink instability, the enhancement of the quasi-adiabatic Speiser ion current causes the current sheet to thin, generating strong density gradients. 
These density gradients result in the rapid growth of the lower-hybrid drift instability (LHDI)~\cite{daughton_nonlinear_2002,daughton_electromagnetic_2003,daughton_nonlinear_2004}, which is localized at the current sheet edges due to the strong stabilization effect of finite $\beta$~\cite{davidson_effects_1977}. 
As demonstrated by Ref.~\onlinecite{daughton_nonlinear_2002} and~\onlinecite{daughton_nonlinear_2004}, the nonlinear evolution of the LHDI induces resonant scattering of crossing Speiser ions into the non-crossing region of phase space. 
This process leads to a net negative charge within the current sheet and the formation of an electrostatic potential across it, which in turn results in a thickening of the current sheet~\cite{sitnov_structure_2006}. 
As the radius of curvature increases due to the thickening of the current sheet, Speiser ions are subsequently scattered back into the hot, isotropic ion population, thereby relaxing the ion VDFs toward equilibrium. 

We show that, to maintain the stability of the current sheet, the ion anisotropy is constrained by two limits set by curvature scattering. 
Nevertheless, as recent theoretical developments and \textit{in situ} observations suggest, plasma anisotropy and agyrotropy play a crucial role in the force balance of magnetotail-like current sheets~\cite{artemyev_configuration_2021,an_configuration_2022,egedal_plasma_2023}. 
Since electrons are generally isotropic, their pressure force accounts for only up to 10\%-30\% of the magnetic tension~\cite{artemyev_contribution_2019}. 
Recent numerical simulations have further revealed that even a small population of Speiser ions can contribute significantly to maintaining force balance through their anisotropic and agyrotropic pressure contribution~\cite{arnold_pic_2023}. 
Therefore, our results indicate that Speiser ions play a crucial role in maintaining the force balance of the current sheet and consistently determine the magnetic field curvature, i.e., the degree of stretching of the current sheet, necessary to ensure its stability.

\section{Conclusions}
We present a new model that describes the limits of ion anisotropy in a magnetotail-like quasi-1D current sheet with three ion populations. 
The anisotropy limits are imposed by the curvature scattering to maintain stability of the current sheet. 
We validate the model against numerical simulations and \textit{in situ} observations. 
Our results demonstrate that the parallel anisotropy is constrained by a modified Firehose-type current sheet instability, which explicitly incorporates the contribution of cold ion populations to the current density through their curvature drift. 
This refinement provides a more complete description of the stability boundary governing anisotropy-driven dynamics in multi-ion populations plasmas. 
In contrast, our results suggest that the perpendicular anisotropy is limited by a two-beam drift-type instability, which acts to regulate the relative contribution of quasi-adiabatic Speiser ions to the total current density. 
Together, these results offer a unified physical framework for understanding how temperature anisotropy, current sheet stability, and ion dynamics are linked in collisionless magnetotail environments.

\begin{acknowledgments}
    We thank the MMS team for data access and support. 
    L.R. acknowledges support from the Royal Swedish Academy of Sciences grant AST2024-0015, the Knut and Alice Wallenberg Foundation (Dnr. 2022.0087), and the Swedish National Space Agency grant 192/20. 
    CN acknowledges support from the Swedish National Space Agency (SNSA) under Grant 2022-00121, and the Research Council of Norway under contract 300865. 
    The PIC-HC simulation was performed on resources provided by Sigma2 -- the National Infrastructure for High-Performance Computing and Data Storage in Norway. 
    X.A.~is supported by NASA grant No.~80NSSC22K1634. 
    X.A.~acknowledges high-performance computing support from Derecho (\url{https://doi.org/10.5065/qx9a-pg09}), provided by NCAR's Computational and Information Systems Laboratory and sponsored by the National Science Foundation \cite{derecho}. 
    A.V.A. and S.R.K. are supported by NASA grants No.~80NSSC23K0658, 80NSSC25K0074, and 80NSSC22K0752. 
\end{acknowledgments}

\bibliography{main}

\end{document}